# Pulse-to-pulse wavelength switchting of diode based fiber laser for multi-color multi-photon imaging

Matthias Eibl[1], Sebastian Karpf[2], Hubertus Hakert[1], Daniel Weng[1], Torben Blömker[1] and Robert Huber[1]

[1]Institut für Biomedizinische Optik, Universität zu Lübeck, Lübeck, Peter-Monnik-Weg 4, 23562 Lübeck, Germany
[2]Department of Electrical Engineering, University of California, Los Angeles, California 90095, USA

## ABSTRACT

We present an entirely fiber based laser source for non-linear imaging with a novel approach for multi-color excitation. The high power output of an actively modulated and amplified picosecond fiber laser at 1064 nm is shifted to longer wavelengths by a combination of four-wave mixing and stimulated Raman scattering. By combining different fiber types and lengths, we control the non-linear wavelength conversion in the delivery fiber itself and can switch between 1064 nm, 1122 nm, and 1186 nm on-the-fly by tuning the pump power of the fiber amplifier and modulate the seed diodes. This is a promising way to enhance the applicability of short pulsed laser diodes for bio-molecular non-linear imaging by reducing the spectral limitations of such sources. In comparison to our previous work [1, 2], we show for the first time two-photon imaging with the shifted wavelengths and we demonstrate pulse-to-pulse switching between the different wavelengths without changing the configuration.

## 1. INTRODUCTION

Non-linear bio-molecular imaging is a growing field with a strong demand for high power lasers. Current laser systems for non-linear imaging such as two-photon excitation fluorescence (TPEF) microscopy or stimulated Raman scattering (SRS) microscopy employ ultra-short pulse lasers (~100 fs-10 ps) in order to achieve the high intensities required for non-linear excitation [3, 4]. Besides high power, the accessible spectral bandwidth, i.e. tunability, is an important characteristic as, for example in TPEF microscopy, many fluorophores with different absorption peaks should be excited. Commonly used Ti:Sa lasers fulfil both requirements. However, especially in combination with additional optical parametric oscillators (OPO), they are expensive, have a large form factor and are not inherently fiber compatible. Hence, we investigate alternative approaches. Several groups are developing laser diode based light sources for non-linear imaging and have shown their great potential [5-9]. Although they usually do not achieve such ultra-short pulses, they have been applied successfully in TPEF imaging [6, 10-13] and it has been shown that the image quality compared to a Ti:Sa is the same if the same duty cycle is applied [1, 14]. A downside is, however, that these lasers often lack spectral flexibility as they operate at one wavelength only. Here, we demonstrate a simple and straightforward method to extend the spectral coverage of these semiconductor and fiber based lasers by a combination of SRS and four-wave-mixing (FWM) in the delivery fiber itself.

## 2. METHODS

The investigated laser scheme is depicted in Fig. 1 A). We use a home-built master oscillator fiber power amplifier (MOFPA) where the output of a narrowband 1064 nm laser diode is actively modulated by a fast electro-optical-modulator (EOM) to adjustable 50 ps- 5 ns pulses. These ~100 mW pulses are amplified in a three stage ytterbium doped fiber amplifier (YDFA) up to kWs of peak power. The 1064 nm laser diode is pre-modulated to prevent depleting of the YDFAs by light leaking through the EOM. The EOM has a bandwidth of 12 GHz and is driven either by a homebuilt pulse generator (PG) (400 ps-5 ns) or by a fast pulse generator (35 ps-250 ps, AlnairLabs EPG-210). The pulse repetition rate can be chosen freely and we typically use rates between 100 kHz-1 MHz. Compared to our previous work [1] we optimized the fiber length by cut-back method in order to enable high output power at different wavelengths from one output fiber port (Fig. 1 B). By a combination of 2 m SMF28 and 2 m Hi1060 fiber after the 4 m long double clad (DC) ytterbium fiber (LMA-YDF-10/125-9M), we achieve non-linear wavelength conversion in the delivery fiber itself and can switch between 1064 nm, 1122 nm, and 1186 nm on-the-fly by the following measures: for maximum 1064 nm output, we tune the output power until the Raman threshold of 1064 nm is reached (cp. Fig. 1 B, blue curve). To maximize 1122 nm light output, we turn on the 1122 nm seed diode and increase the power of the DC-YDFA and, thereby, the 1122 nm power, until the onset of a spectral component at 1186 nm is visible. If the power of the DC-YDFA gets increased even more, one would expect a broad Raman gain around 1180 nm representing the Raman spectrum of fused silica (cp. Fig. 1 B, grey area below red curve). However, we observed a narrow spectral line at 1186 nm. As the energy difference between these three spectral components is the same, we believe that 1186 nm photons are generated by four-wave mixing (FWM). However, as there is no phase matching between either of these three wavelengths, FWM cannot be the dominant process for generating high peak power pulses at 1186 nm. We suggest that a few photons generated by FWM seed the second Raman order and, therefore, a combination of SRS and FWM leads to high power and spectrally narrowband pulses at 1186 nm. To maximize this 1186 nm output, we turn again the 1122 nm seed diode on and increase the pump power of the DC-YDFA until the power gets shifted to the next Raman band at 1257 nm.

It can be seen that the maximum power of the three wavelength components is determined by the initial peak pulse power of the 1064 nm seed, by the on/off status of the 1122 nm seed diode, and the amplification in the DC-YDFA. By modulating the 1064 nm and 1122 nm seed power combined with appropriate pump power for the DC-YDFA, shifting between either two of these three wavelength is possible in a pulse-to-pulse manner.

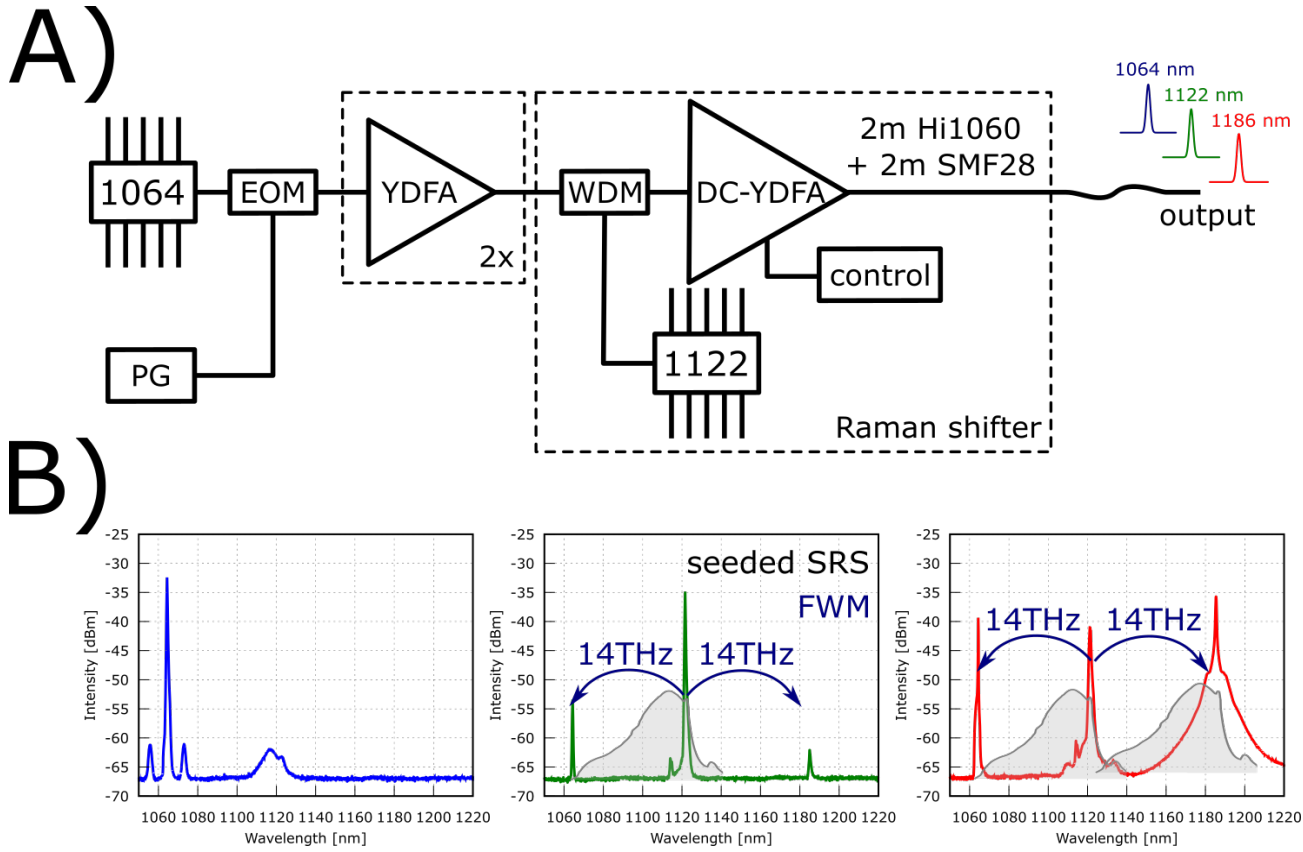

Fig. 1: A) fiber MOPA setup. The output of a spectrally narrowband 1064 nm seed laser diode gets modulated with a fast EOM to 50 ps – 1 ns pulses. These pulses are amplified in a multistage YDFA up to 1-2 kW of peak power. The Raman gain of the delivery fiber is used to shift the 1064 nm output to longer wavelengths. B) Output spectra of the laser when optimized for different wavelengths.

## 3. RESULTS

### 3.1 TPEF imaging with wavelength shifted light source

First, we show the applicability of our wavelength shifting scheme for TPEF imaging of red fluorophores. The imaging setup is described elsewhere [15, 16]. Figure 2 shows TPEF images of labeld COS-7 cells. Mitochondria are labeled with Atto-594 and tubulin with STAR-635. Atto-594 shows a strong two-photon absorbance at 1122 nm and at 1186 nm and STAR-635 has a strong absorbance at 1186 nm but hardly at 1122 nm. With this in mind, we imaged the cells once with the laser tuned to 1122 nm and once with 1186 nm output. By subtracting the intensity images of both acquisitions, we could identify the tubulin structure very well (cp. Fig. 2 left) and by multiplying both images, we could highlight the mitochondria (cp. Fig. 2 middle). Both images overlayed show very well that both structures are clearly distinguishable only by different excitation wavelengths.

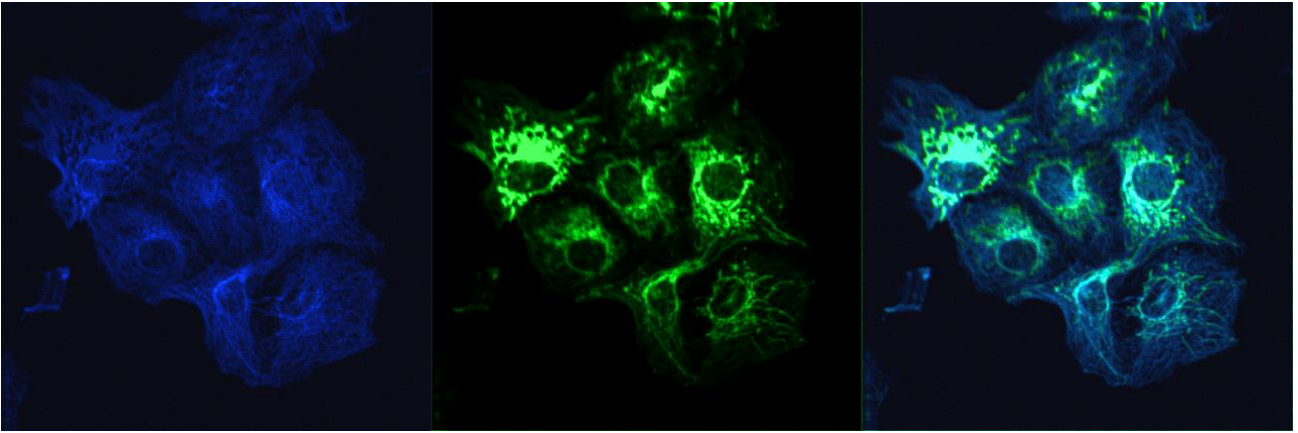

Fig. 2: TPEF imaging of labeld COS-7 cells with 1122 nm and 1186 nm pulses from a single fiber output of a MOPA laser. Mitochondria labeled with Atto-594 are colored green and tubulin labeled with STAR-635 are colored blue. Left image: difference between 1122 nm and 1186 nm excited TPEF intensity image. Middle image: Multiplication of 1122 nm and 1186 nm excited TPEF intensity image. Right image: overlay of both images.

### 3.2 Pulse-to-pulse wavelength shifting

Furthermore, we wanted to show if it is possible to switch between these wavelengths rapidly, i.e. from pulse to pulse. As mentioned above, the key parameters to change the wavelengths are the initial 1064 nm seed power, the 1122 nm seed diode, and the pump power for the DC-YDFA.

First, switching between 1064 nm and 1122 nm is shown. The ouput power at 1064 nm was increased until the Raman threshold was reached. Then, the 1122 nm seed diode was turned on for every second pulse. We modulated the 1122 nm seed diode to ~40 ns pulses to minimize depletion of the following DC-YDFA. The temporal pulse output can be seen in figure 3, left. The 1 ns long pulses had a repetition rate of 50 kHz and were recorded with a homemade 500 MHz freespace photodiode connected to a 1 GHz realtime oscilloscope. The two spectral components were separated with a long-pass 1100 nm (green curve) and a short-pass 1100 nm (blue curve) filter. Both components show similar pulse heights and the peak power was ~600 W. Conversion efficiency from 1064 nm to 1122 nm was better than 90%. The right graph in figure 3 shows the spectral output. Both components have a spectral 3dB width below 50 pm.

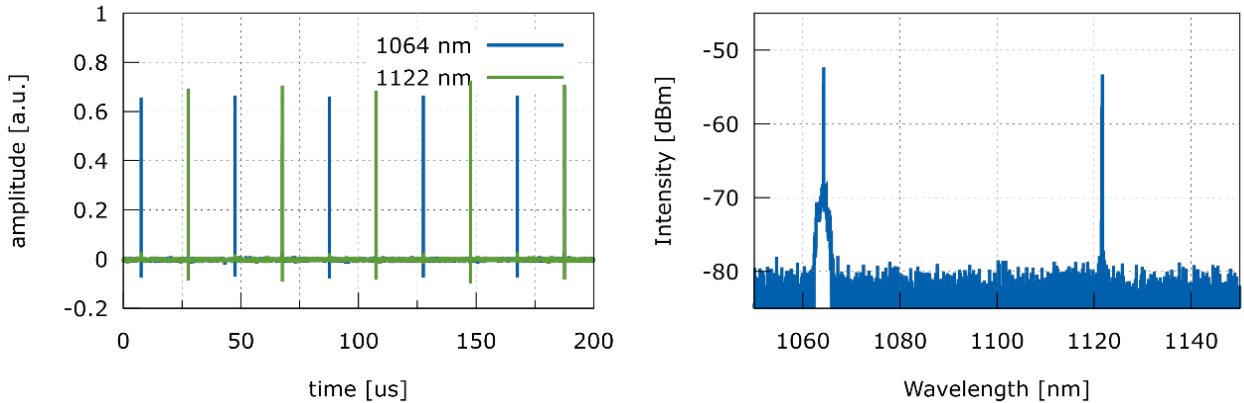

Fig. 3: Pulse-to-pulse switching between 1064 nm and 1122 nm. left: oscilloscope traces; right: output spectrum.

Second, we present wavelength switching between 1122 nm and 1186 nm. To efficiently generate 1122 nm, the seed laser diode for this wavelength has to be turned on. Hence, turning on and off this laser diode cannot be used to switch between the two longer wavelengths. To shift from 1122 nm to 1186 nm the initial 1064 nm peak power has to be increased. We achieve this by modulating the 1064 nm seed diode. The DC-YDFA pump power is increased until 1186 nm pulses are optimized. Then, the 1064 nm seed power is reduced for every second pulse so that the peak power is not sufficient to generate 1186 nm light. As it can be seen in figure 4 (left), the pulses have similar power levels. The two spectral components were separated with a longpass 1150 nm filter and a shortpass 1150 nm. Also, we verified that less than 1% of the pulse amplitude is remaining 1064 nm with a laser line filter at 1064 nm (not shown). Less than 10% of the other wavelength was in the respective pulse power. Figure 4 right, shows the spectral output. The 1186 nm line has a spectral width of less than 50 pm

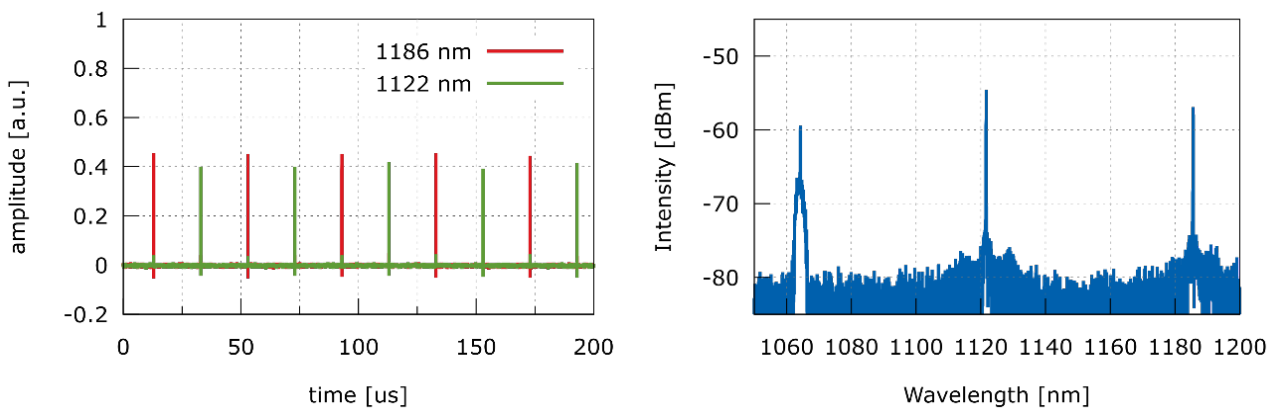

Fig. 4: Pulse-to-pulse switching between 1122 nm and 1186 nm. left: oscilloscope traces; right: output spectrum.

Finally, we present wavelength switching from 1064 nm to 1186 nm from pulse-to-pulse. To achieve this, the 1064 nm seed diode modulation depth was increased so that the lower energy pulses remained below the Raman threshold for 1064 nm and the high energy pulses, together with 1122 nm seed, have enough power to generate the 1186 nm component. As there is a higher peak power necessary to generate 1186 nm light, the overall peak power is higher for that component than for the 1064 nm pulses (cp. fig. 5, left).

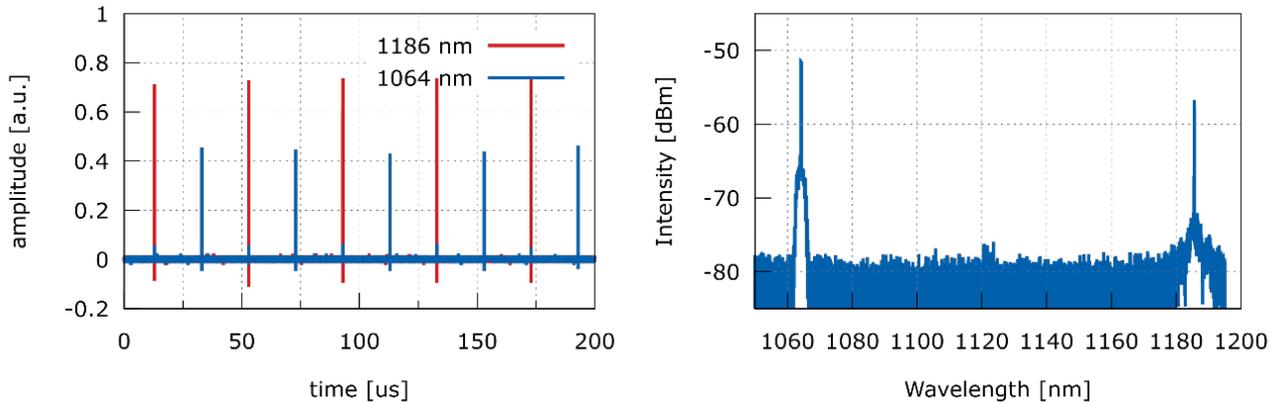

Fig. 5: Pulse-to-pulse switching between 1064 nm and 1186 nm. left: oscilloscope traces; right: output spectrum.

## 4. CONCLUSION AND OUTLOOK

We present a fiber amplified diode laser which can be shifted in wavelength by electronically controlling the MOPA's seed , a second seed diode, and the pump power of the last fiber amplifier. We presented excitation of red fluorophores with the two Raman shifted longer wavelengths at 1122 nm and 1186 nm. Furthermore, we presented pulse-to-pulse wavelength switching from 1064 nm to 1122 nm, from 1122 nm to 1186 nm, and from 1064 nm to 1186 nm. In all configurations, the spectral output remained narrowband and most of the power was in the desired spectral component. With the presented novel multi-color approach it is possible to implement various new wavelengths to make laser diode based sources more flexible for two photon imaging. Also, the all single mode fiber based approach and longer pulses which are less sensitive to pulse break-up in fibers are ideal for future endoscopic setups. In future experiments, we would like to show a combination with fast OCT [17-21] from a single fiber output endoscope.